\documentstyle[prd,aps,preprint,tighten,epsfig]{revtex}

\begin{document}

\draft

\title{Distinguishable RGE running effects between
Dirac neutrinos and Majorana neutrinos with vanishing Majorana
CP-violating phases}

\author{{\bf Zhi-zhong Xing}\thanks{E-mail: xingzz@mail.ihep.ac.cn}
~ and ~ {\bf He Zhang}\thanks{E-mail: zhanghe@mail.ihep.ac.cn}}
\address{CCAST (World Laboratory), P.O. Box 8730,
Beijing 100080, China \\
and Institute of High Energy Physics,
Chinese Academy of Sciences, \\
P.O. Box 918, Beijing 100049, China}
\maketitle

\begin{abstract}
In a novel parametrization of neutrino mixing and in the
approximation of $\tau$-lepton dominance, we show that the
one-loop renormalization-group equations (RGEs) of Dirac neutrinos
are different from those of Majorana neutrinos even if two
Majorana CP-violating phases vanish. As the latter can keep
vanishing from the electroweak scale to the typical seesaw scale,
it makes sense to distinguish between the RGE running effects of
neutrino mixing parameters in Dirac and Majorana cases. The
differences are found to be quite large in the minimal
supersymmetric standard model with sizable $\tan\beta$, provided
the masses of three neutrinos are nearly degenerate or have an
inverted hierarchy.
\end{abstract}


\section{Introduction}

Since 1998, a number of successful neutrino oscillation
experiments \cite{SK,SNO,KM,CHOOZ,K2K} have provided us with very
convincing evidence that neutrinos are massive and lepton flavors
are mixed. A global fit of current experimental data
\cite{Vissani} yields two neutrino mass-squared differences:
$\Delta m^{2}_{21} \equiv m^2_2 - m^2_1 =(8.0 \pm 0.3) \times
10^{-5}~ {\rm eV^2}$ and $\Delta m^{2}_{32} \equiv m^2_3 - m^2_2 =
\pm (2.5 \pm 0.3) \times 10^{-3} ~{\rm eV^2}$. The upper bound of
every neutrino mass is expected to be $m^{}_i < 0.23 ~{\rm eV}$
(for $i=1,2,3$) \cite{Mohapatra}. Three neutrino mixing angles,
associated respectively with solar, atmospheric and CHOOZ
experiments, are well restricted: $30^\circ \lesssim
\theta^{}_{12} \lesssim 38^\circ$, $36^\circ \lesssim
\theta^{}_{23} \lesssim 54^\circ$ and $\theta^{}_{13} <10^\circ$
at the $99\%$ confidence level \cite{Vissani}. But whether
neutrinos are Dirac or Majorana particles remains an open
question. If the neutrinoless double-beta ($\beta\beta^{}_{0\nu}$)
decay is eventually observed, we shall make sure that neutrinos
are Majorana particles. If there is no experimental signal for the
$\beta\beta^{}_{0\nu}$ decay, however, we shall be unable to
conclude that neutrinos are just Dirac particles \cite{Review}.
Although most theorists believe that neutrinos should be Majorana
fermions, there {\it do} exist some interesting models which treat
massive neutrinos as Dirac fermions \cite{Dirac}. Before the
nature of neutrinos is ultimately identified by the future
neutrino experiments, it is worthwhile to study the phenomenology
of both Dirac and Majorana neutrinos.

Since many viable neutrino models are proposed at a superhigh
energy scale, one has to take account of radiative corrections to
their consequences on lepton flavor mixing and neutrino
oscillations at low energy scales. Inversely, one may examine the
evolution of those observed neutrino mixing parameters from the
electroweak scale $\Lambda^{}_{\rm EW} \sim 10^2$ GeV up to a
superhigh energy scale by using their renormalization-group
equations (RGEs). This kind of study is very useful, both for
model building itself and for distinguishing different models. It
is also useful to reveal the intrinsic differences between Dirac
and Majorana neutrinos, as we shall explicitly show in this paper.

A simple extension of the standard model (SM) or the minimal
supersymmetric standard model (MSSM) is to introduce very heavy
right-handed Majorana neutrinos and keep the Lagrangian of
electroweak interactions invariant under $\rm SU(2)^{}_L \times
U(1)^{}_Y$ gauge transformation. The smallness of left-handed
Majorana neutrino masses can then be explained via the well-known
seesaw mechanism \cite{SS}. Below the typical seesaw scale
$\Lambda^{}_{\rm SS} \sim 10^{14}$ GeV, where heavy Majorana
neutrinos become decoupled, the effective neutrino coupling matrix
$\kappa$ obeys
\begin{equation}
16\pi^2 \frac{{\rm d}\kappa}{{\rm d}t} = \alpha^{}_{\rm M} \kappa
+ C \left [ \left (Y^{}_lY^\dagger_l \right ) \kappa + \kappa
\left (Y^{}_l Y^\dagger_l \right )^T \right ] \;
\end{equation}
at the one-loop level \cite{RGE}, where $t\equiv \ln
(\mu/\Lambda_{\rm SS})$ with $\mu$ being an arbitrary
renormalization scale between $\Lambda_{\rm EW}$ and $\Lambda_{\rm
SS}$, $Y^{}_l$ is the charged-lepton Yukawa coupling matrix,
$C=-1.5$ (SM) or $C=1$ (MSSM), $\alpha^{}_{\rm M} \approx -3g^2_2
+ 6 y^2_t + \lambda$ (SM) or $\alpha^{}_{\rm M} \approx -1.2g^2_1
-6g^2_2 + 6 y^2_t$ (MSSM). Here $g^{}_1$ and $g^{}_2$ are the
gauge couplings, $y^{}_t$ stands for the top-quark Yukawa
coupling, and $\lambda$ denotes the Higgs self-coupling in the SM.

If neutrinos are Dirac particles, their Yukawa coupling matrix
$Y^{}_\nu$ must be extremely suppressed in magnitude to reproduce
the light neutrino masses of ${\cal O}(1)$ eV or smaller at low
energy scales. The running of $Y^{}_\nu$ from a superhigh energy
scale (e.g., $\Lambda^{}_{\rm SS}$) down to $\Lambda_{\rm EW}$ is
governed by the one-loop RGE \cite{Xing05}
\begin{equation}
16\pi^2 \frac{{\rm d} \omega}{{\rm d}t} = 2 \alpha^{}_{\rm D}
\omega + C \left [ \left (Y^{}_lY^\dagger_l \right ) \omega +
\omega \left (Y^{}_l Y^\dagger_l \right )\right ] \; ,
\end{equation}
where $\omega \equiv Y^{}_\nu Y^\dagger_\nu$, $\alpha^{}_{\rm D}
\approx -0.45g^2_1 - 2.25g^2_2 + 3y^2_t$ (SM) or $\alpha^{}_{\rm
D} \approx -0.6g^2_1 - 3g^2_2 + 3y^2_t$ (MSSM). In writing out Eq.
(2), we have safely neglected those tiny terms of ${\cal
O}(\omega^2)$.

One may use Eq. (1) or (2) to derive the explicit RGEs for
neutrino mixing parameters in the flavor basis where $Y^{}_l$ is
diagonal and real (positive). In this basis, we have $\kappa =
{\cal V}^{}_{\rm M} \overline{\kappa} {\cal V}^T_{\rm M}$ with
$\overline{\kappa} = {\rm Diag}\{\kappa^{}_1, \kappa^{}_2,
\kappa^{}_3\}$ for Majorana neutrinos; or $\omega = {\cal
V}^{}_{\rm D} \overline{\omega} {\cal V}^\dagger_{\rm D}$ with
$\overline{\omega} = {\rm Diag} \{ y^2_1, y^2_2, y^2_3 \}$ for
Dirac neutrinos. ${\cal V}^{}_{\rm M}$ or ${\cal V}_{\rm D}$ is
just the lepton flavor mixing matrix. The unitarity violation of
${\cal V}^{}_{\rm M}$ is extremely tiny and can safely be
neglected in all realistic seesaw models \cite{Zhou}. At
$\Lambda_{\rm EW}$, Majorana neutrino masses are $m^{}_i = v^2
\kappa^{}_i$ (SM) or $m^{}_i = v^2 \kappa^{}_i \sin^2\beta$ (MSSM)
and Dirac neutrino masses are $m^{}_i = v y^{}_i$ (SM) or $m^{}_i
= v y^{}_i \sin\beta$ (MSSM) with $v \approx 174$ GeV. A general
parametrization of ${\cal V}^{}_{\rm M}$ or ${\cal V}_{\rm D}$ is
${\cal V}^{}_{\rm M} = Q^{}_{\rm M} U P^{}_{\rm M}$ or ${\cal
V}^{}_{\rm D} = Q^{}_{\rm D}UP^{}_{\rm D}$, where $P^{}_{\rm M}$
(or $P^{}_{\rm D}$) and $Q^{}_{\rm M}$ (or $Q^{}_{\rm D}$) are two
diagonal phase matrices, and $U$ is a unitary matrix containing
three mixing angles and one CP-violating phase. Note that
$P^{}_{\rm D}$, $Q^{}_{\rm D}$ and $Q^{}_{\rm M}$ have no physical
significance, but $P^{}_{\rm M} = {\rm Diag} \{ e^{i\rho},
e^{i\sigma}, 1\}$ consists of two non-trivial (physical) phases
which are commonly referred to as the Majorana CP-violating
phases. A novel parametrization of $U$ is \cite{FX97}
\begin{eqnarray}
U & = & \left ( \matrix{ s^{}_l s^{}_{\nu} c + c^{}_l c^{}_{\nu}
e^{-i\phi} & s^{}_l c^{}_{\nu} c - c^{}_l s^{}_{\nu} e^{-i\phi} &
s^{}_l s \cr c^{}_l s^{}_{\nu} c - s^{}_l c^{}_{\nu} e^{-i\phi} &
c^{}_l c^{}_{\nu} c + s^{}_l s^{}_{\nu} e^{-i\phi} & c^{}_l s \cr
- s^{}_{\nu} s   & - c^{}_{\nu} s   & c \cr } \right ) \; ,
\end{eqnarray}
where $c^{}_l \equiv \cos\theta^{}_l$, $s^{}_{\nu} \equiv
\sin\theta^{}_{\nu}$, $c \equiv \cos\theta$, and so on. This
parametrization, together with the approximation of $\tau$-lepton
dominance (i.e., $Y^{}_l Y^\dagger_l \approx {\rm Diag} \{ 0, 0,
y^2_\tau\}$ in view of $y^2_e \ll y^2_\mu \ll y^2_\tau$), allows
us to obtain the following RGEs for two Majorana phases $\rho$ and
$\sigma$ \cite{Xing05}:
\begin{eqnarray}
\dot{\rho} & = & \frac{C y^2_\tau}{16\pi^2} \left [
\widehat{\zeta}^{}_{12} c^2_\nu s^2 c^{}_{(\sigma -\rho)}
s^{}_{(\sigma -\rho)} + \widehat{\zeta}^{}_{13} \left (s^2_\nu s^2
- c^2 \right ) c^{}_\rho s^{}_\rho + \widehat{\zeta}^{}_{23}
c^2_\nu s^2 c^{}_\sigma s^{}_\sigma \right ] \; ,
\nonumber \\
\dot{\sigma} & = & \frac{C y^2_\tau}{16\pi^2} \left [
\widehat{\zeta}^{}_{12} s^2_\nu s^2 c^{}_{(\sigma -\rho)}
s^{}_{(\sigma -\rho)} + \widehat{\zeta}^{}_{13} s^2_\nu s^2
c^{}_\rho s^{}_\rho + \widehat{\zeta}^{}_{23} \left ( c^2_\nu s^2
- c^2 \right ) c^{}_\sigma s^{}_\sigma \right ] \; ,
\end{eqnarray}
where $\dot{\rho} \equiv {\rm d}{\rho}/{\rm d}t$, $\dot{\sigma}
\equiv {\rm d}{\rho}/{\rm d}t$, $\widehat{\zeta}^{}_{ij} \equiv 4
\kappa^{}_i \kappa^{}_j/\left ( \kappa^2_i - \kappa^2_j \right )$,
$c^{}_a \equiv \cos a$ and $s^{}_a \equiv \sin a$ (for $a = \rho$,
$\sigma$ or $\sigma -\rho$). A particularly interesting feature of
Eq. (4) is that $\rho = \sigma = 0$ at a specific energy scale
leads to $\dot{\rho} =\dot{\sigma} =0$, implying that $\rho$ and
$\sigma$ can keep vanishing at any energy scales between
$\Lambda^{}_{\rm EW}$ and $\Lambda^{}_{\rm SS}$. In this case,
only three mixing angles $(\theta^{}_l, \theta^{}_\nu, \theta)$
and the so-called Dirac CP-violating phase $\phi$ undergo the RGE
evolution. Note that a kind of underlying flavor symmetry may
actually forbid two Majorana phases to take non-zero values in a
concrete neutrino model. It is therefore meaningful to ask whether
the RGE running behaviors of Majorana neutrinos with $\rho =
\sigma = 0$ are identical to those of Dirac neutrinos. The purpose
of this paper is just to answer such a question.

In section II, we show that the one-loop RGEs of Majorana
neutrinos with $\rho = \sigma = 0$ are analytically similar to
those of Dirac neutrinos, but their expressions are not exactly
identical. Section III is devoted to a numerical analysis of the
RGE running behaviors of three mixing angles and the Dirac
CP-violating phase, and to a careful comparison between the cases
of Dirac and Majorana neutrinos. Four different neutrino mass
spectra are taken into account in our calculations. We find that
the differences between Dirac and Majorana neutrinos in their RGE
running effects can be quite large in the MSSM with sizable
$\tan\beta$, provided the masses of three neutrinos are nearly
degenerate or have an inverted hierarchy. A brief summary of the
main results is given in section IV.

\section{RGEs of Dirac and Majorana neutrinos}

The one-loop RGEs for three Yukawa coupling eigenvalues of Dirac
neutrinos ($y^{}_i$ with $i=1,2,3$) and their four flavor mixing
parameters ($\theta^{}_l$, $\theta^{}_\nu$, $\theta$ and $\phi$)
have been derived in Ref. \cite{Xing05}. Here we replace $y^{}_i$
by $m^{}_i$. The RGEs of three neutrino masses, three mixing
angles and one CP-violating phase can then be written as
\begin{eqnarray}
\dot{m}^{}_1 & = & \frac{m^{}_1}{16\pi^2} \left ( \alpha^{}_{\rm
D} + C y^2_\tau s^2_\nu s^2 \right ) \; ,
\nonumber \\
\dot{m}^{}_2 & = & \frac{m^{}_2}{16\pi^2} \left ( \alpha^{}_{\rm
D} + C y^2_\tau c^2_\nu s^2 \right ) \; ,
\nonumber \\
\dot{m}^{}_3 & = & \frac{m^{}_3}{16\pi^2} \left ( \alpha^{}_{\rm
D} + C y^2_\tau c^2 \right ) \; ;
\end{eqnarray}
and
\begin{eqnarray}
\dot{\theta}^{}_l & = & + \frac{Cy^2_\tau}{8\pi^2} ~ c^{}_\nu
s^{}_\nu c c^{}_\phi \frac{m^2_3 \left ( m^2_2 - m^2_1 \right
)}{\left (m^2_3 - m^2_1 \right ) \left ( m^2_3 - m^2_2 \right )}
\; ,
\nonumber \\
\dot{\theta}^{}_\nu & = & - \frac{Cy^2_\tau}{16\pi^2} ~ c^{}_\nu
s^{}_\nu \left [ s^2 \frac{m^2_2 + m^2_1}{m^2_2 - m^2_1} - c^2
\frac{2m^2_3 \left ( m^2_2 - m^2_1 \right )}{\left ( m^2_3 - m^2_1
\right ) \left (m^2_3 - m^2_2 \right )} \right ] \; ,
\nonumber \\
\dot{\theta} \; & = & - \frac{Cy^2_\tau}{16\pi^2} ~ c s \left (
s^2_\nu \frac{m^2_3 + m^2_1}{m^2_3 - m^2_1} + c^2_\nu \frac{m^2_3
+ m^2_2}{m^2_3 - m^2_2} \right ) \; ,
\nonumber \\
\dot{\phi} \; & = & - \frac{Cy^2_\tau}{8\pi^2} \left ( c^2_l -
s^2_l \right ) c^{-1}_l s^{-1}_l c^{}_\nu s^{}_\nu c s^{}_\phi
\frac{m^2_3 \left (m^2_2 - m ^2_1 \right )}{\left (m^2_3 - m^2_1
\right ) \left (m^2_3 - m^2_2 \right )} \; ,
\end{eqnarray}
where $c^{}_\phi \equiv \cos\phi$ and $s^{}_\phi \equiv \sin\phi$.
Note that the neutrino mass-squared differences $m^2_3 - m^2_1
\equiv \Delta m^2_{31}$ and $m^2_3 - m^2_2 \equiv \Delta m^2_{32}$
are much larger in magnitude than $m^2_2 - m^2_1 \equiv \Delta
m^2_{21}$, as indicated by current experimental data. Typically,
$\Delta m^2_{21} \approx 8.0 \times 10^{-5} ~{\rm eV}^2$ and
$|\Delta m^2_{31}| \approx |\Delta m^2_{32}| \approx 2.5 \times
10^{-3} ~{\rm eV}^2$ \cite{Vissani}. Among three neutrino mixing
angles, the RGE running of $\theta^{}_\nu$ is expected to be most
significant. The CP-violating phase $\phi$ may significantly
evolve from one energy scale to another, if $\theta^{}_l$ takes
sufficiently small values. These qualitative features will become
clearer in our subsequent numerical calculations.

The one-loop RGEs for three effective coupling eigenvalues of
Majorana neutrinos ($\kappa^{}_i$ with $i=1,2,3$) and their six
flavor mixing parameters ($\theta^{}_l$, $\theta^{}_\nu$,
$\theta$, $\phi$, $\rho$ and $\sigma$) can be found in Ref.
\cite{Xing05}. Here we replace $\kappa^{}_i$ by $m^{}_i$ and take
$\rho = \sigma =0$ at either $\Lambda^{}_{\rm EW}$ or
$\Lambda^{}_{\rm SS}$. As pointed out in section I, Eq. (4)
assures two Majorana phase $\rho$ and $\sigma$ to keep vanishing
at any energy scales between $\Lambda^{}_{\rm EW}$ and
$\Lambda^{}_{\rm SS}$. One may safely simplify the RGEs of
$\theta^{}_l$, $\theta^{}_\nu$, $\theta$ and $\phi$ obtained in
Ref. \cite{Xing05} for Majorana neutrinos by setting $\rho =
\sigma =0$, and then compare them with their Dirac counterparts on
the same footing. In this case, we arrive at
\begin{eqnarray}
\dot{m}^{}_1 & = & \frac{m^{}_1}{16\pi^2} \left ( \alpha^{}_{\rm
M} + 2 C y^2_\tau s^2_\nu s^2 \right ) \; ,
\nonumber \\
\dot{m}^{}_2 & = & \frac{m^{}_2}{16\pi^2} \left ( \alpha^{}_{\rm
M} + 2 C y^2_\tau c^2_\nu s^2 \right ) \; ,
\nonumber \\
\dot{m}^{}_3 & = & \frac{m^{}_3}{16\pi^2} \left ( \alpha^{}_{\rm
M} + 2 C y^2_\tau c^2 \right ) \; ;
\end{eqnarray}
and
\begin{eqnarray}
\dot{\theta}^{}_l & = & + \frac{Cy^2_\tau}{8\pi^2} ~ c^{}_\nu
s^{}_\nu c c^{}_\phi \frac{m^{}_3(m^{}_2 - m^{}_1)}{\left ( m^{}_3
- m^{}_1 \right ) \left (m^{}_3 - m^{}_2 \right )} \; ,
\nonumber \\
\dot{\theta}^{}_\nu & = & - \frac{Cy^2_\tau}{16\pi^2} ~ c^{}_\nu
s^{}_\nu \left [ s^2 \frac{m^{}_2 + m^{}_1}{m^{}_2 - m^{}_1} - c^2
\frac{2 m^{}_3 \left (m^{}_2 - m^{}_1 \right )}{\left (m^{}_3 -
m^{}_1 \right ) \left (m^{}_3 - m^{}_2 \right )} \right ] \; ,
\nonumber \\
\dot{\theta} \; & = & - \frac{Cy^2_\tau}{16\pi^2} ~ c s \left (
s^2_\nu \frac{m^{}_3 + m^{}_1}{m^{}_3 - m^{}_1} + c^2_\nu
\frac{m^{}_3 + m^{}_2}{m^{}_3 - m^{}_2} \right ) \; ,
\nonumber \\
\dot{\phi} \; & = & - \frac{Cy^2_\tau}{8\pi^2} \left ( c^2_l -
s^2_l \right ) c^{-1}_l s^{-1}_l c^{}_\nu s^{}_\nu c s^{}_\phi
\frac{m^{}_3 \left (m^{}_2 - m^{}_1 \right )}{\left (m^{}_3 -
m^{}_1 \right ) \left (m^{}_3 - m^{}_2 \right )} \; .
\end{eqnarray}
As a consequence of $\Delta m^2_{21} \ll |\Delta m^2_{31}| \approx
|\Delta m^2_{32}|$, the mixing angle $\theta^{}_\nu$ is most
sensitive to radiative corrections. The RGE evolution of the
CP-violating phase $\phi$ depends strongly on the smallness of
$\theta^{}_l$, on the other hand. These qualitative features are
essentially analogous to what we have pointed out for Dirac
neutrinos.

It is interesting to note that Eq. (7) can actually be obtained
from Eq. (5) with the replacements $\alpha^{}_{\rm D}
\Longrightarrow \alpha^{}_{\rm M}$ and $C \Longrightarrow 2C$,
while Eq. (8) can be achieved from Eq. (6) with the replacements
$m^2_i \Longrightarrow m^{}_i$ (for $i=1,2,3$). These similarities
and differences imply that it is very non-trivial to distinguish
between the RGE running behaviors of Dirac neutrinos and Majorana
neutrinos with vanishing Majorana CP-violating phases.

A rephasing-invariant description of leptonic CP violation in
neutrino oscillations is to make use of the Jarlskog parameter
$\cal J$ \cite{J}. It explicitly reads as ${\cal J} = c^{}_l
s^{}_l c^{}_\nu s^{}_\nu c s^2 s^{}_\phi$ in the parametrization
taken in Eq. (3) for either Dirac or Majorana neutrinos. The
one-loop RGE of $\cal J$ has been derived in Ref. \cite{Xing05}.
Taking $\rho =\sigma =0$ in the Majorana case, we obtain a
simplified expression of $\dot{\cal J}^{\rm M}$, \small
\begin{equation}
\dot{\cal J}^{\rm M} \; =\; \frac{Cy^2_\tau}{16\pi^2} ~ {\cal
J}^{\rm M} \left [ \left (c^2_\nu - s^2_\nu \right ) s^2
\frac{m^{}_2 + m^{}_1}{m^{}_2 - m^{}_1} + \left ( c^2 - s^2_\nu
s^2 \right ) \frac{m^{}_3 + m^{}_1}{m^{}_3 - m^{}_1} + \left ( c^2
- c^2_\nu s^2 \right ) \frac{m^{}_3 + m^{}_2}{m^{}_3 - m^{}_2}
\right ] \; ,
\end{equation}
\normalsize which is very analogous to $\dot{\cal J}^{\rm D}$ of
Dirac neutrinos, \small
\begin{equation}
\dot{\cal J}^{\rm D} \; =\; \frac{Cy^2_\tau}{16\pi^2} ~ {\cal
J}^{\rm D} \left [ \left (c^2_\nu - s^2_\nu \right ) s^2
\frac{m^2_2 + m^2_1}{m^2_2 - m^2_1} + \left ( c^2 - s^2_\nu s^2
\right ) \frac{m^2_3 + m^2_1}{m^2_3 - m^2_1} + \left ( c^2 -
c^2_\nu s^2 \right ) \frac{m^2_3 + m^2_2}{m^2_3 - m^2_2} \right ]
\; .
\end{equation}
\normalsize
It is obvious that Eq. (10) can be obtained from Eq.
(9) with the replacements $m^{}_i \Longrightarrow m^2_i$ (for
$i=1,2,3$). Note that $\dot{\cal J}^{\rm D} \propto {\cal J}^{\rm
D}$ (or $\dot{\cal J}^{\rm M} \propto {\cal J}^{\rm M}$) holds.
This result implies that the Jarlskog parameter will keep
vanishing at any energy scales between $\Lambda^{}_{\rm EW}$ and
$\Lambda^{}_{\rm SS}$, if it initially vanishes at either
$\Lambda^{}_{\rm EW}$ or $\Lambda^{}_{\rm SS}$.

Two comments are in order. First, three mixing angles
$\theta^{}_l$, $\theta^{}_\nu$ and $\theta$ in our parametrization
can simply be related to $\theta^{}_{12}$, $\theta^{}_{23}$ and
$\theta^{}_{13}$ in the standard parametrization advocated by the
Particle Data Group \cite{PDG}. The relations are \cite{FX06}
\begin{equation}
\theta^{}_{12} \approx \theta^{}_\nu \; , ~~~~~~ \theta^{}_{23}
\approx \theta \; , ~~~~~~ \theta^{}_{13} \approx \theta^{}_l
\sin\theta \;
\end{equation}
in the leading-order approximation. Thus $30^\circ \lesssim
\theta^{}_\nu \lesssim 38^\circ$, $36^\circ \lesssim \theta
\lesssim 54^\circ$ and $\theta^{}_l < 17^\circ$ are expected to
hold, as indicated by the global fit done in Ref. \cite{Vissani}.
Second, the approximation of $\tau$-lepton dominance used in our
analytical calculations is very reliable. In other words, the
contributions of $y^2_e$ and $y^2_\mu$ to all of the RGEs listed
above are negligibly small. This observation has been verified by
our numerical calculations, in which there is no special
assumption or approximation.

\section{Numerical illustration and comparison}

In view of the fact that the absolute mass scale of three light
neutrinos and the sign of $\Delta m^{2}_{32}$ remain unknown at
present, let us consider four typical patterns of the neutrino
mass spectrum:
\begin{itemize}
\item Normal hierarchy (NH): $m^{}_1 \ll m^{}_2 \ll m^{}_3$. For
simplicity, we typically take $m^{}_1 = 0$ at $\Lambda^{}_{\rm
EW}$ in our numerical calculations. Then $m^{}_2=\sqrt{\Delta
m^2_{21}}$ and $m^{}_3 = \sqrt{|\Delta m^2_{32}| + \Delta
m^2_{21}}$ can be determined from current experimental data.

\item Inverted hierarchy (IH): $m^{}_3 \ll m^{}_1 \lesssim m^{}_2
$. For simplicity, w typically take $m^{}_3=0$ at $\Lambda^{}_{\rm
EW}$ in our numerical calculations. Then $m^{}_2=\sqrt{|\Delta
m^2_{32}|}$ and $m^{}_1=\sqrt{|\Delta m^2_{32}| - \Delta
m^2_{21}}$ can be determined from current experimental data.

\item Near degeneracy (ND) with $\Delta m^2_{32} >0$: $m^{}_1
\lesssim m^{}_2 \lesssim m^{}_3$. For simplicity, we typically
take $m^{}_1 = 0.2~{\rm eV}$ at $\Lambda^{}_{\rm EW}$ in our
numerical calculations.

\item Near degeneracy (ND) with $\Delta m^2_{32} <0$: $m^{}_3
\lesssim m^{}_1 \lesssim m^{}_2$. For simplicity, we typically
take $m^{}_1 = 0.2~{\rm eV}$ at $\Lambda^{}_{\rm EW}$ in our
numerical calculations.
\end{itemize}
In addition, we take $\Delta m^2_{21} = 8.0 \times 10^{-5} ~ {\rm
eV}^2$, $|\Delta m^2_{32}| = 2.5 \times 10^{-3} ~ {\rm eV}^2$,
$\theta^{}_\nu = 33.8^\circ$, $\theta = 45^\circ$, $\theta^{}_l =
0.5^\circ$ and $\phi = 90^\circ$ as typical inputs at
$\Lambda^{}_{\rm EW} \sim M^{}_Z$ in our numerical calculations.

\subsection{Neutrino masses}

In either the SM or the MSSM with small $\tan\beta$, the RGE
running behaviors of three neutrino masses are dominated by
$\alpha^{}_{\rm D}$ or $\alpha^{}_{\rm M}$. The
$y^2_\tau$-associated term of $\dot{m}^2_i$ in Eq. (5) or (7)
becomes important only when $\tan\beta$ takes sufficiently large
values in the MSSM \cite{Lindner}. Note that $\alpha^{}_{\rm M} =
2 \alpha^{}_{\rm D}$ holds in the MSSM, in which the running
effects of $m^{}_i$ for Majorana neutrinos are twice as large as
those for Dirac neutrinos.

The first plot in Fig. 1 illustrates the ratios $R_i \equiv
m^{}_i(\mu)/m^{}_i(M^{}_Z)$ changing with the energy scale $\mu$
in the SM for Dirac and Majorana neutrinos, where $m^{}_1(M^{}_Z)
= 0.2$ eV and $M^{}_H = 180$ GeV (the Higgs mass) have typically
been input. Since the running of $m^{}_i$ is governed by
$\alpha^{}_{\rm D}$ or $\alpha^{}_{\rm M}$, $R^{}_1 \approx R^{}_2
\approx R^{}_3$ holds to a high degree of accuracy. Furthermore,
the behaviors of $R^{}_i$ are actually independent of the initial
value of $m^{}_1$ and possible patterns of the neutrino mass
spectrum. We observe that $R^{}_i$ in the Majorana case is always
larger than $R^{}_i$ in the Dirac case, and their discrepancy can
be as large as 0.7 at $\mu \sim \Lambda^{}_{\rm SS} \sim 10^{14}$
GeV.

The relation $R^{}_1 \approx R^{}_2 \approx R^{}_3$ is also a very
good approximation in the MSSM with small $\tan\beta$, as shown by
the second plot in Fig. 1, where $\tan\beta = 10$ has been input.
It is clear that $R^{}_i$ in the Dirac case is numerically
distinguishable from $R^{}_i$ in the Majorana case, in particular
when the energy scale $\mu$ far exceeds $M^{}_Z$.

If $\tan\beta$ is sufficiently large, the common scaling of three
neutrino masses in the RGE evolution will fail \cite{Lindner}. The
splitting of $R^{}_1$, $R^{}_2$ and $R^{}_3$, which increases with
the energy scale $\mu$, is illustrated by the third plot in Fig. 1
with the input $\tan\beta = 50$. One can see that $R^{}_i$ in the
Dirac case is always smaller than $R^{}_i$ in the Majorana case,
and their discrepancy is distinguishable at the scales $\mu \gg
M^{}_Z$.

\subsection{Neutrino mixing parameters}

Radiative corrections to three neutrino mixing angles, the Dirac
CP-violating phase and the Jarlskog parameter are all controlled
by the $\tau$-lepton Yukawa coupling eigenvalue $y^{}_\tau$.
Because of $y^2_\tau/(8\pi^2) \approx 1.3 \times 10^{-6}$ (SM) or
$y^2_\tau/(8\pi^2) \approx 1.3 \times 10^{-6} \left ( 1 +
\tan^2\beta \right )$ (MSSM) at $M^{}_Z$, significant RGE running
effects are expected to appear in the MSSM case when $\tan\beta$
is sufficiently large \cite{Haba}. To illustrate, here we simply
concentrate on the MSSM with $\tan\beta=50$ and consider four
typical patterns of the neutrino mass spectrum in our subsequent
discussions and calculations.

(1) In the NH case with $m^{}_1 =0$, the RGEs of $\theta^{}_l$,
$\theta^{}_\nu$, $\theta$, $\phi$ and $\cal J$ can be simplified
as
\begin{eqnarray}
\dot{\theta}^{}_l &  = & + \frac{Cy^2_\tau}{16\pi^2} ~ c^{}_\nu
s^{}_\nu c c^{}_\phi r \ , \ \ \ \ \ \ \ \ \dot{\theta}^{}_\nu = -
\frac{Cy^2_\tau}{16\pi^2} ~ c^{}_\nu s^{}_\nu \left (s^2 - c^2 r
\right ) \ ,
\nonumber \\
\dot{\theta} \; & = & - \frac{Cy^2_\tau}{16\pi^2} ~ c s \left (
1+r \right ) \; , \ \ \ \ \ \ \ \dot{\phi} \; = -
\frac{Cy^2_\tau}{16\pi^2} \left ( c^2_l - s^2_l \right ) c^{-1}_l
s^{-1}_l c^{}_\nu s^{}_\nu c s^{}_\phi r \; ,
\nonumber \\
\dot{\cal J} \; & =  & \; \frac{Cy^2_\tau}{16\pi^2} ~ {\cal J}
\left [ 2 \left (s^2_\nu s^2 - c^2 \right ) - \left ( c^2 -
c^2_\nu s^2 \right ) r \right ] \; ,
\end{eqnarray}
in which $r = 2 m^2_2/(m^2_3 - m^2_2 )$ for Dirac neutrinos or $r
= 2m^{}_2/(m^{}_3 - m^{}_2 )$ for Majorana neutrinos. Current
experimental data yield $r^{}_{\rm D} \approx 0.06$ and $r^{}_{\rm
M} \approx 0.4$. Both of them are too small to compensate for the
strong suppression induced by $y^2_\tau$ in Eq. (12). Thus the RGE
corrections to those flavor mixing and CP-violating parameters are
not important in the NH case. Note, however, that the radiative
correction to $\phi$ can be very significant when $\theta^{}_l$ is
extremely small or becomes vanishing. We find that $\phi$ quickly
approaches its quasi-fixed point $\phi^{}_{\rm QF} = 0$ or $\pi$
in the $\theta^{}_l \rightarrow 0$ limit, an interesting
phenomenon which is remarkably different from the non-trivial
quasi-fixed point of $\phi$ discovered in the general ($\rho \neq
\sigma \neq 0$) case for Majorana neutrinos \cite{Luo}. One can
also see that both ${\cal J} = 0$ and $\dot{\cal J} = 0$ hold when
$\theta^{}_l$ vanishes; i.e., CP is a good symmetry in this limit.

(2) In the IH case with $m^{}_3 =0$, we arrive at
\begin{eqnarray}
\dot{\theta}^{}_l &  = & \dot{\phi}  = 0, \ \ \ \ \
\dot{\theta}^{}_\nu  = - \frac{Cy^2_\tau}{16\pi^2} ~ c^{}_\nu
s^{}_\nu s^2 r^\prime \ , \ \ \ \ \ \dot{\theta}   =
+\frac{Cy^2_\tau}{16\pi^2} ~ c s \; ,
\nonumber \\
\dot{\cal J} \; & = & \; \frac{Cy^2_\tau}{16\pi^2} ~ {\cal J}
\left [ 3c^2 - 1  + \left ( s^2_\nu - c^2_\nu \right ) s^2
r^\prime \right ] \; ,
\end{eqnarray}
where $r^\prime
 = ( m^2_2 + m^2_1 )/(m^2_2 - m^2_1 )$ for Dirac
neutrinos or $r^\prime
 = ( m^{}_2 + m^{}_1 )/(m^{}_2 - m^{}_1 )$ for
Majorana neutrinos. We observe that radiative corrections to
$\theta^{}_l$ and $\phi$ are vanishingly small, and the correction
to $\theta$ is also insignificant. Nevertheless, the RGE running
effects of $\theta^{}_\nu$ and $\cal J$ may get enhanced by
$r^\prime$, whose typical value reads $r^\prime_{\rm D} \approx
60$ or $r^\prime_{\rm M} \approx 120$ at $M^{}_Z$. Fig. 2
illustrates the evolution of $\theta^{}_\nu$ and $\cal J$ in the
IH case. The discrepancy between Dirac and Majorana cases is
obviously distinguishable for both parameters, when the energy
scale is much higher than $M^{}_Z$. In particular, ${\cal J}^{\rm
D} \sim 2 {\cal J}^{\rm M}$ holds at $\mu \sim \Lambda^{}_{\rm
SS}$, because the corresponding value of $\theta^{}_\nu$ for
Majorana neutrinos is roughly half of that for Dirac neutrinos.

(3) In the ND case with $\Delta m^2_{32} >0$ and $m^{}_1 = 0.2$
eV, the RGE corrections to those neutrino mixing parameters can
significantly be enhanced by the ratios $( m^2_i + m^2_j )/( m^2_i
- m^2_j )$ in Eqs. (6) and (10) for Dirac neutrinos, or by the
ratios $( m^{}_i + m^{}_j )/( m^{}_i - m^{}_j )$ in Eqs. (8) and
(9) for Majorana neutrinos. We illustrate the typical evolution
behaviors of $\theta^{}_l$, $\theta^{}_\nu$, $\theta$ and $\phi$
in Fig. 3. One can see that Majorana neutrinos undergo the RGE
corrections more significantly than Dirac neutrinos. The
discrepancy between these two cases is about $10^\circ$ for either
$\theta$ or $\phi$ at $\mu \gg M^{}_Z$. It is therefore possible
to distinguish the running of Majorana neutrinos from that of
Dirac neutrinos. The difference between ${\cal J}^{\rm D}$ and
${\cal J}^{\rm M}$ is insignificant even at $\mu \sim
\Lambda^{}_{\rm SS}$, as shown in Fig. 4, partly because the
increase (or decrease) of $\theta^{}_l$ can somehow compensate for
the decrease (or increase) of $\theta$ and $\phi$ in the Majorana
(or Dirac) case.

(4) In the ND case with $\Delta m^2_{32} <0$ and $m^{}_1 = 0.2$
eV, we get similar enhancements in the RGEs of those neutrino
mixing parameters induced by the ratios $( m^2_i + m^2_j )/( m^2_i
- m^2_j )$ for Dirac neutrinos, or by $( m^{}_i + m^{}_j )/(
m^{}_i - m^{}_j )$ for Majorana neutrinos. However, only the
running of $\theta$ is sensitive to the sign flip of $\Delta
m^2_{32}$, as one can see from Eqs. (6) and (8)--(10), in which
$\dot{\theta}^{}_\nu$ and $\dot{\cal J}$ are dominated by the term
proportional to $(m^2_2 + m^2_1)/(m^2_2 - m^2_1)$ (Dirac) or
$(m^{}_2 + m^{}_1)/(m^{}_2 - m^{}_1)$ (Majorana). Then the
numerical results for $\theta^{}_l$, $\theta^{}_\nu$, $\phi$ and
${\cal J}$ in the present case are very similar to those in the ND
case with $\Delta m^2_{32} >0$. For simplicity, we only illustrate
the evolution of $\theta$ in Fig. 5 by taking $\Delta m^2_{32}
<0$. It is obvious that the running behavior of $\theta$ for
either Dirac or Majorana neutrinos in Fig. 5 is essentially
opposite (or complementary) to that in Fig. 3, just due to the
sign flip of $\Delta m^2_{32}$.

\section{Summary}

The main goal of this paper is to examine whether the RGE running
behaviors of Majorana neutrinos are still different from those of
Dirac neutrinos, if two Majorana CP-violating phases vanish at a
given energy scale. For this purpose, it is essential to choose a
suitable parametrization of the $3\times 3$ lepton flavor mixing
matrix, such that its two Majorana phases keep vanishing in the
RGE evolution from one scale to another. We have pointed out that
the novel parametrization used in Ref. \cite{Xing05}, which
consists of the mixing angles $(\theta^{}_l, \theta^{}_\nu,
\theta)$ and the CP-violating phases $(\phi, \rho, \sigma)$, {\it
does} fulfill this requirement. Taking $\rho = \sigma =0$ at the
electroweak scale, we have carefully compared the similarities and
differences between the RGEs of $\theta^{}_l$, $\theta^{}_\nu$,
$\theta$ and $\phi$ for Majorana neutrinos and those for Dirac
neutrinos. Our numerical calculations show that it is possible to
distinguish between these two cases in the MSSM with sizable
$\tan\beta$, in particular when the masses of three neutrinos are
nearly degenerate or have an inverted hierarchy.

Of course, the numerical examples presented in this work are
mainly for the purpose of illustration. The point is that the
nature of neutrinos determines their RGE running behaviors, and
the latter may be crucial for building a realistic neutrino model.
We expect that our analysis can not only complement those previous
studies of radiative corrections to the physical parameters of
Dirac and Majorana neutrinos, but also help us understand the
dynamical role of Majorana phases in a more general picture of
flavor physics.

\acknowledgments{This work is supported in part by the National
Nature Science Foundation of China.}

\newpage

\begin{figure}
\vspace{3.5cm}
\epsfig{file=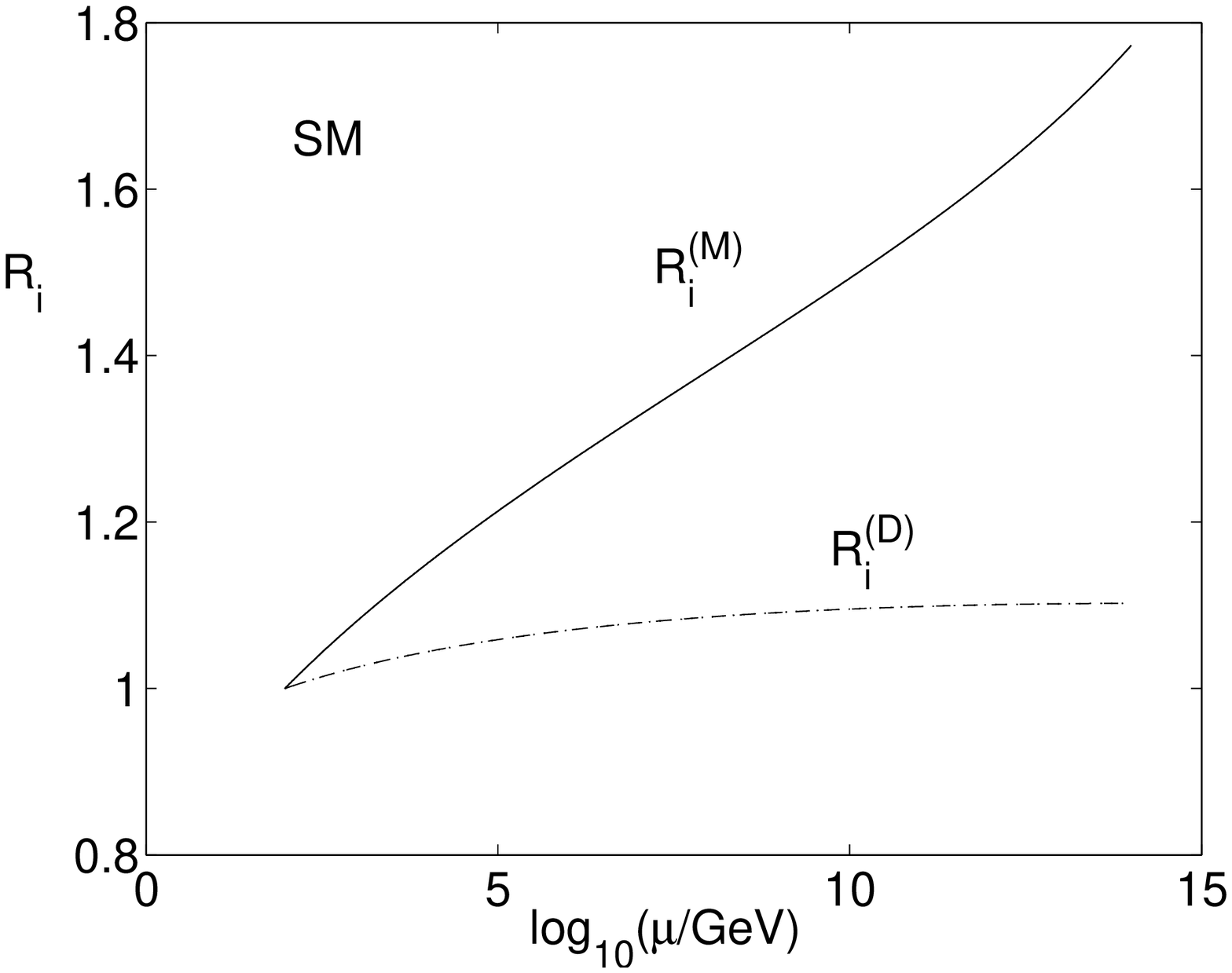,bbllx=-7cm,bblly=9cm,bburx=-3cm,bbury=13cm,%
width=1.8cm,height=1.8cm,angle=0,clip=0}\vspace{4.1cm}
\end{figure}
\begin{figure}
\epsfig{file=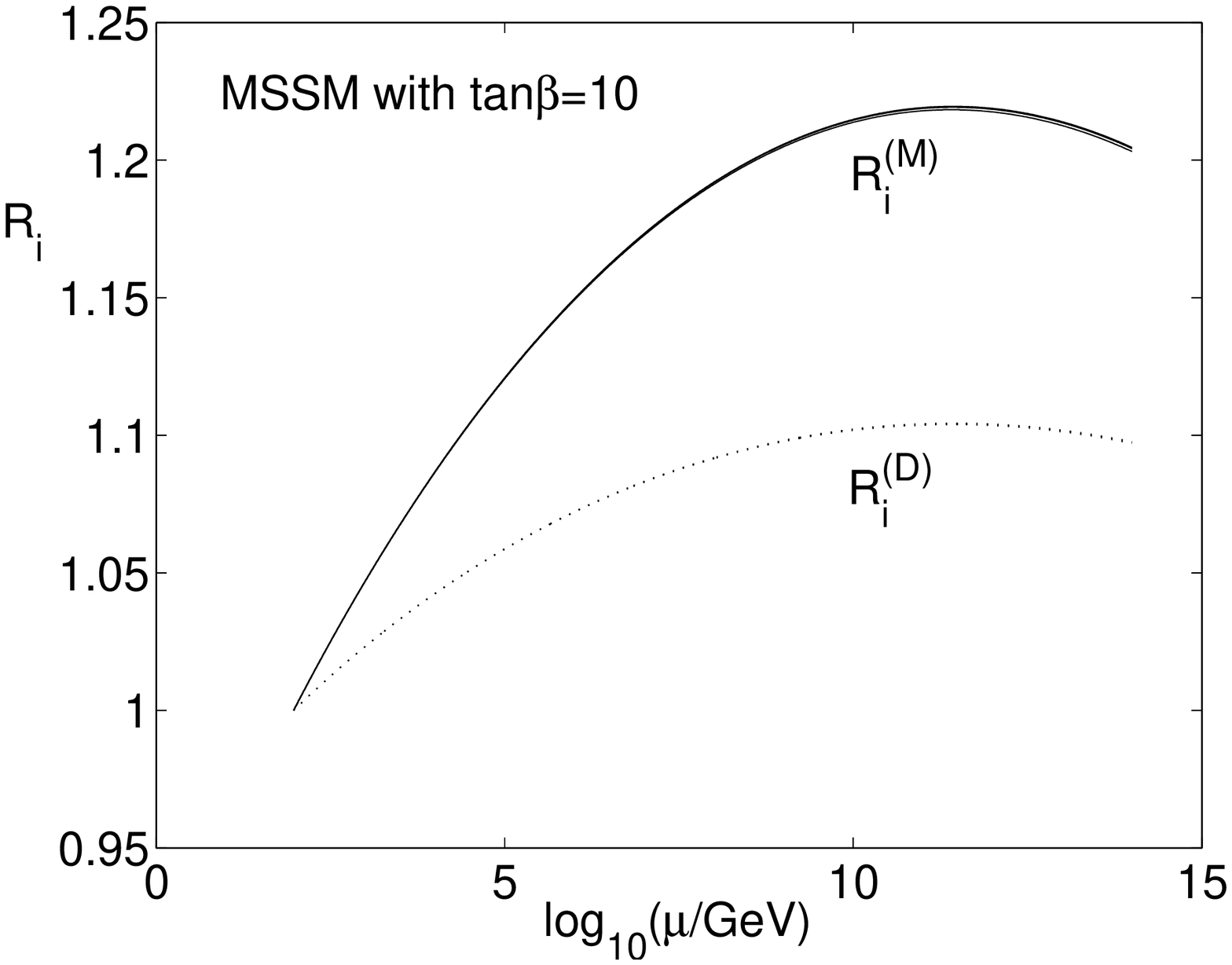,bbllx=-7cm,bblly=8.8cm,bburx=-3cm,bbury=12.8cm,%
width=1.8cm,height=1.8cm,angle=0,clip=0}\vspace{4cm}
\end{figure}
\begin{figure}
\epsfig{file=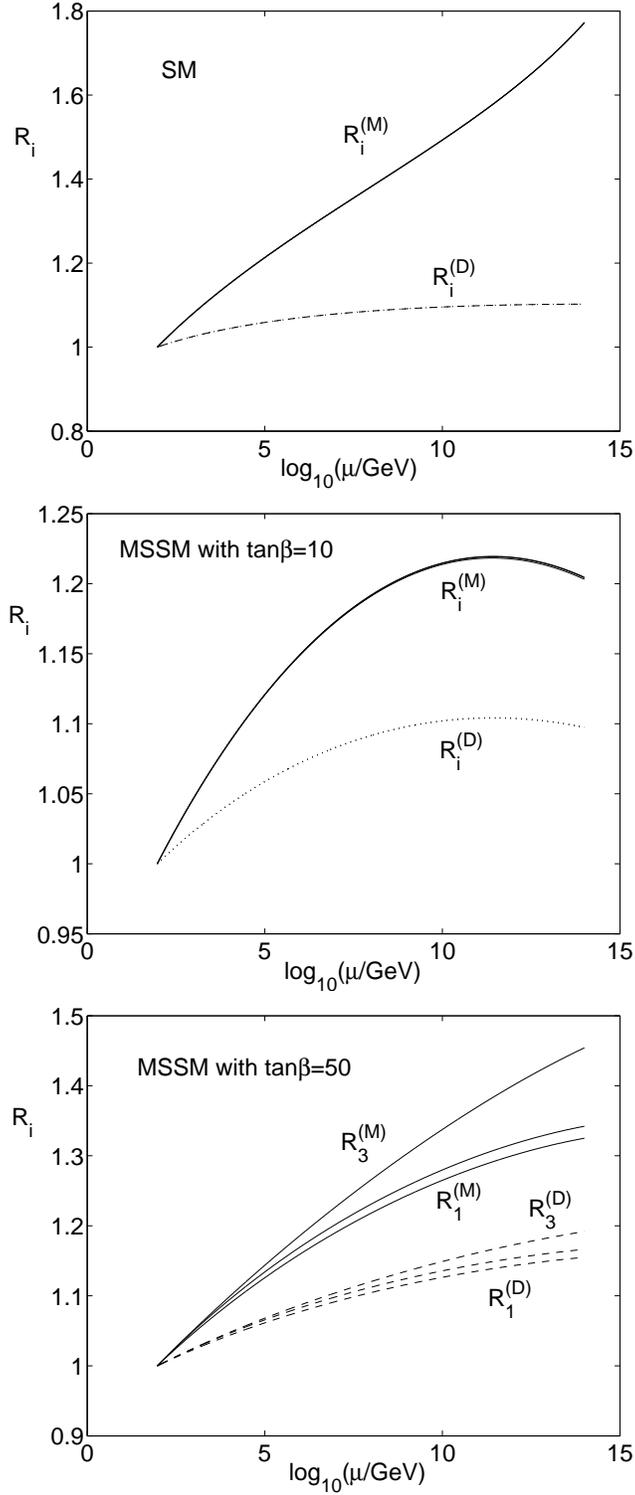,bbllx=-7cm,bblly=8.8cm,bburx=-3cm,bbury=12.8cm,%
width=1.8cm,height=1.8cm,angle=0,clip=0}\vspace{1.65cm}
\caption{The running neutrino mass ratios
$R^{}_{i}=m^{}_{i}(\mu)/m^{}_{i}(M^{}_Z)$ (for $i=1,2,3$), where
the dashed and solid curves stand respectively for the Dirac and
Majorana cases.}\vspace{6cm}
\end{figure}

\begin{figure}[t]
\vspace{-0.5cm}
\epsfig{file=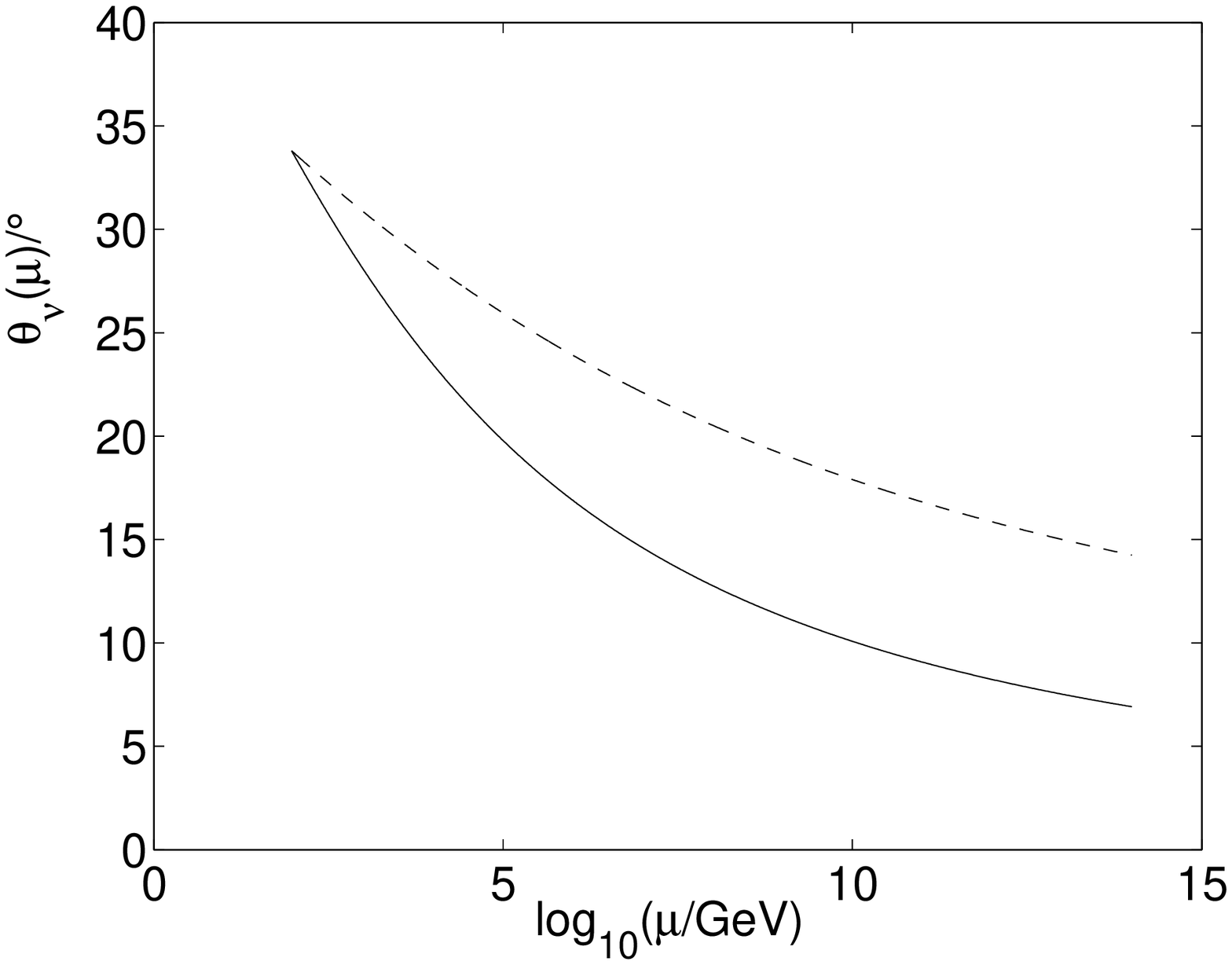,bbllx=1.5cm,bblly=18.5cm,bburx=5.5cm,bbury=22.5cm,%
width=1.5cm,height=1.5cm,angle=0,clip=0}\vspace{-2.3cm}
\end{figure}
\begin{figure}
\epsfig{file=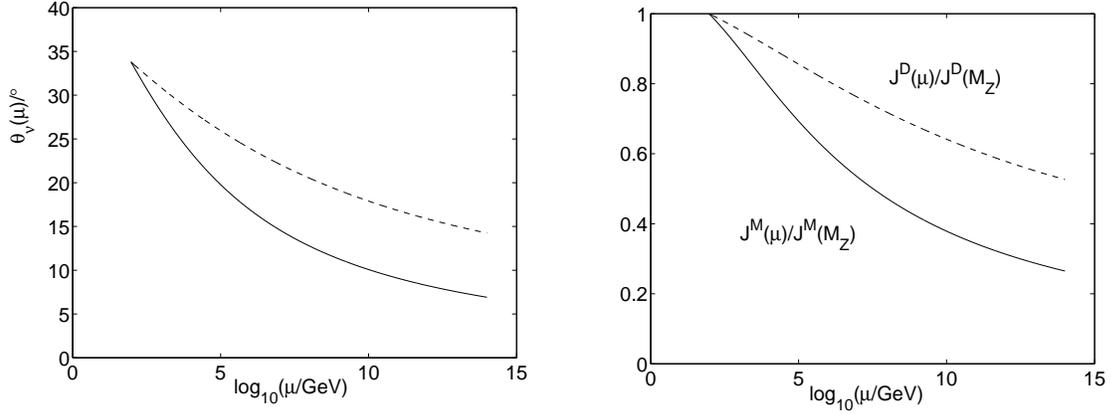,bbllx=-19cm,bblly=18.5cm,bburx=-15cm,bbury=22.5cm,%
width=1.5cm,height=1.5cm,angle=0,clip=0}\vspace{5.1cm}
\caption{The running behaviors of $\theta^{}_\nu$ and ${\cal J}$
in the IH case with $\tan\beta = 50$ and $m^{}_3 = 0$ at $M^{}_Z$
within the MSSM, where the dashed and solid curves stand
respectively for the Dirac and Majorana cases, and ${\cal J}^{\rm
D} (M^{}_Z) = {\cal J}^{\rm M} (M^{}_Z) \approx 0.0014$.}
\end{figure}

\begin{figure}[t]
\vspace{-0.5cm}
\epsfig{file=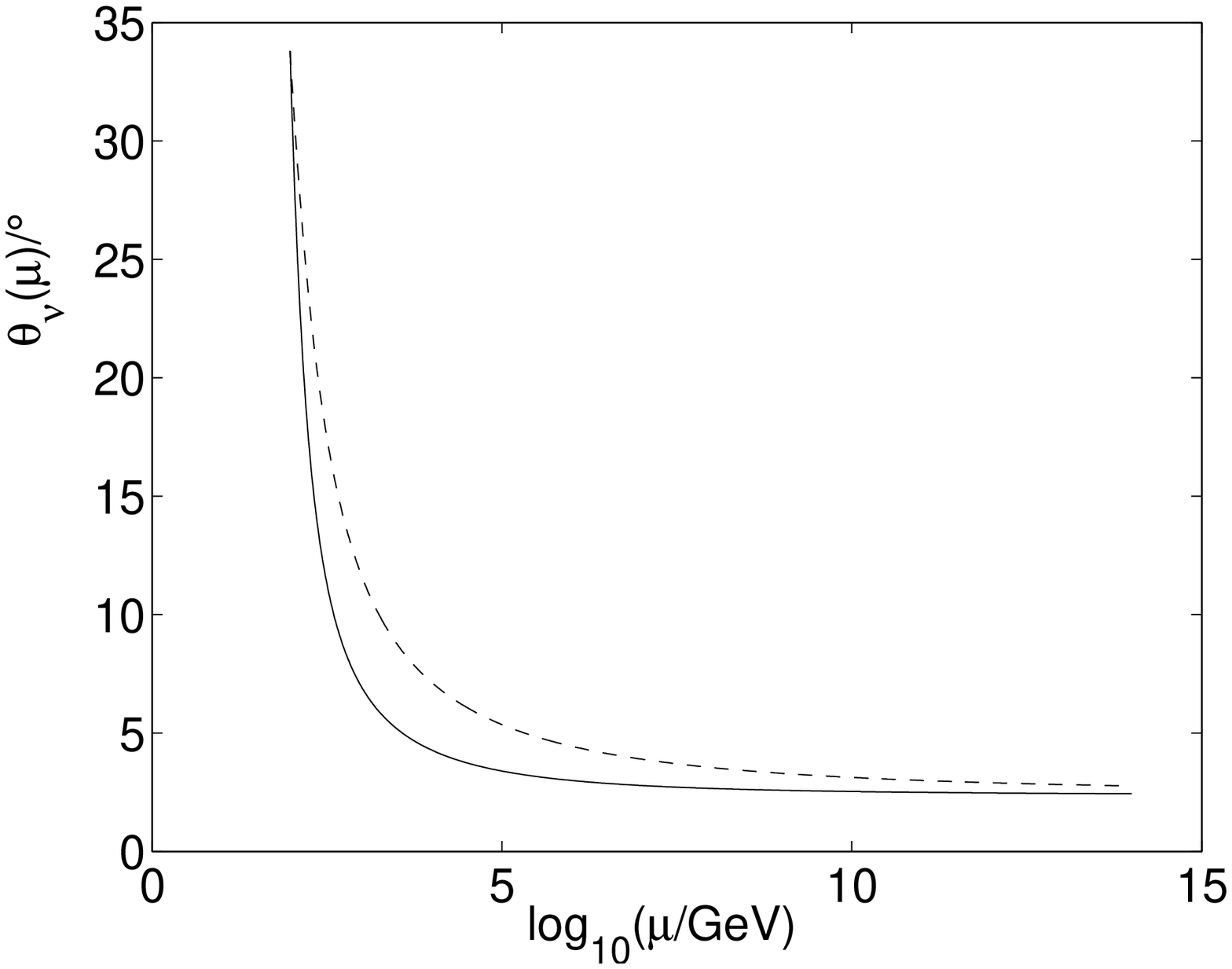,bbllx=1.5cm,bblly=18.5cm,bburx=5.5cm,bbury=22.5cm,%
width=1.5cm,height=1.5cm,angle=0,clip=0}\vspace{3.5cm}
\end{figure}
\begin{figure}
\epsfig{file=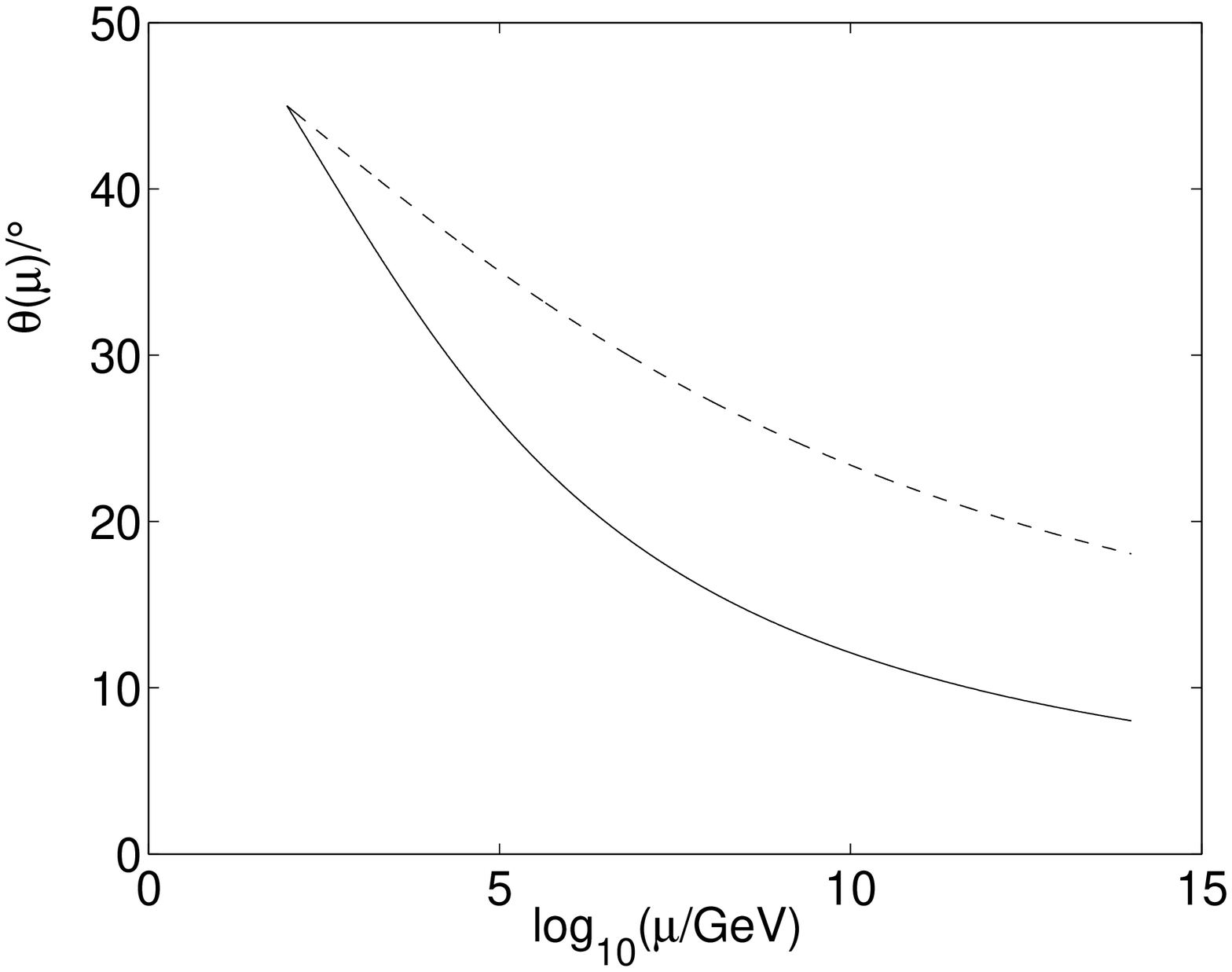,bbllx=1.5cm,bblly=18.5cm,bburx=5.5cm,bbury=22.5cm,%
width=1.5cm,height=1.5cm,angle=0,clip=0}\vspace{-8.2cm}
\end{figure}
\begin{figure}
\epsfig{file=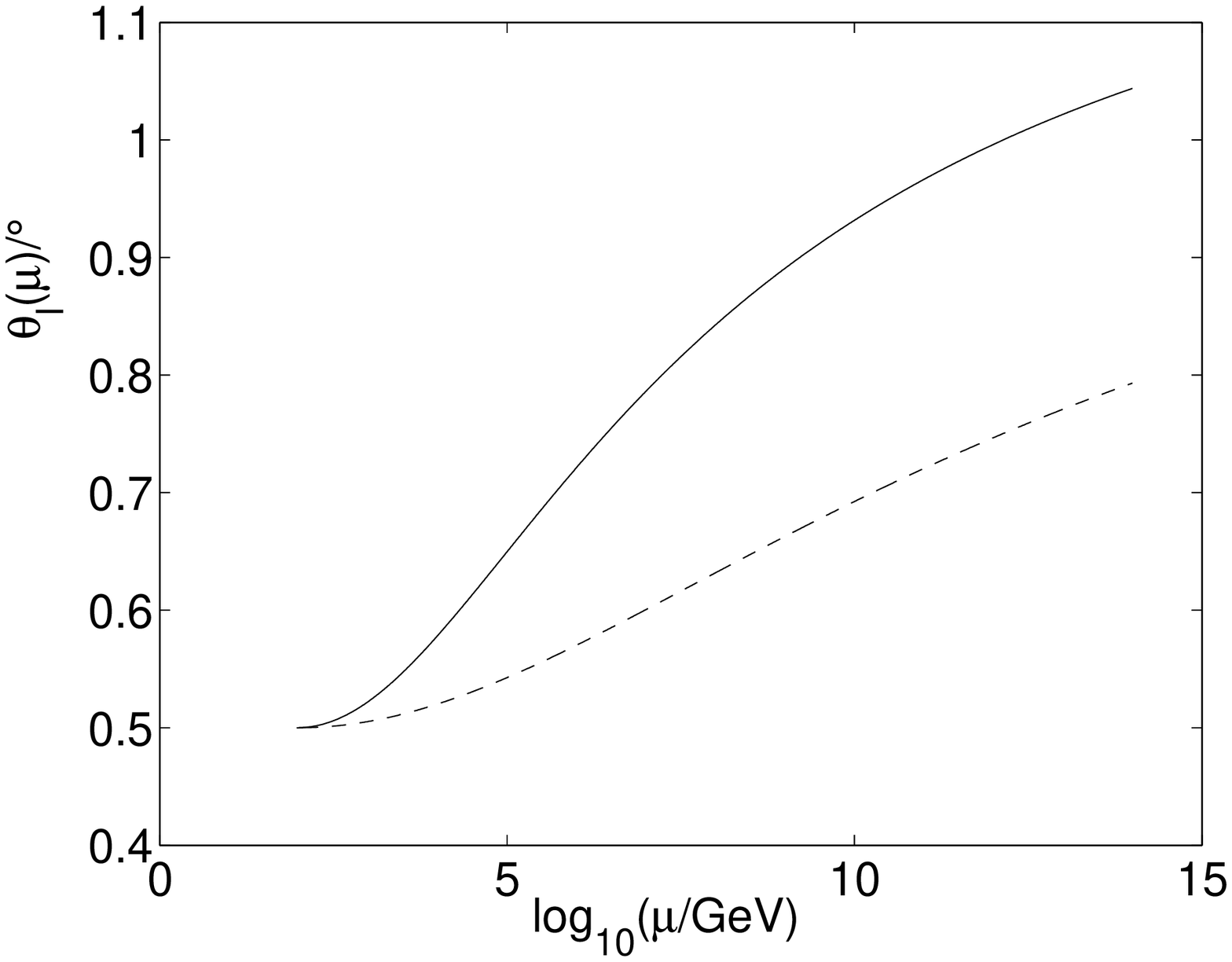,bbllx=-19cm,bblly=18.5cm,bburx=-15cm,bbury=22.5cm,%
width=1.5cm,height=1.5cm,angle=0,clip=0}\vspace{3.3cm}
\end{figure}
\begin{figure}
\epsfig{file=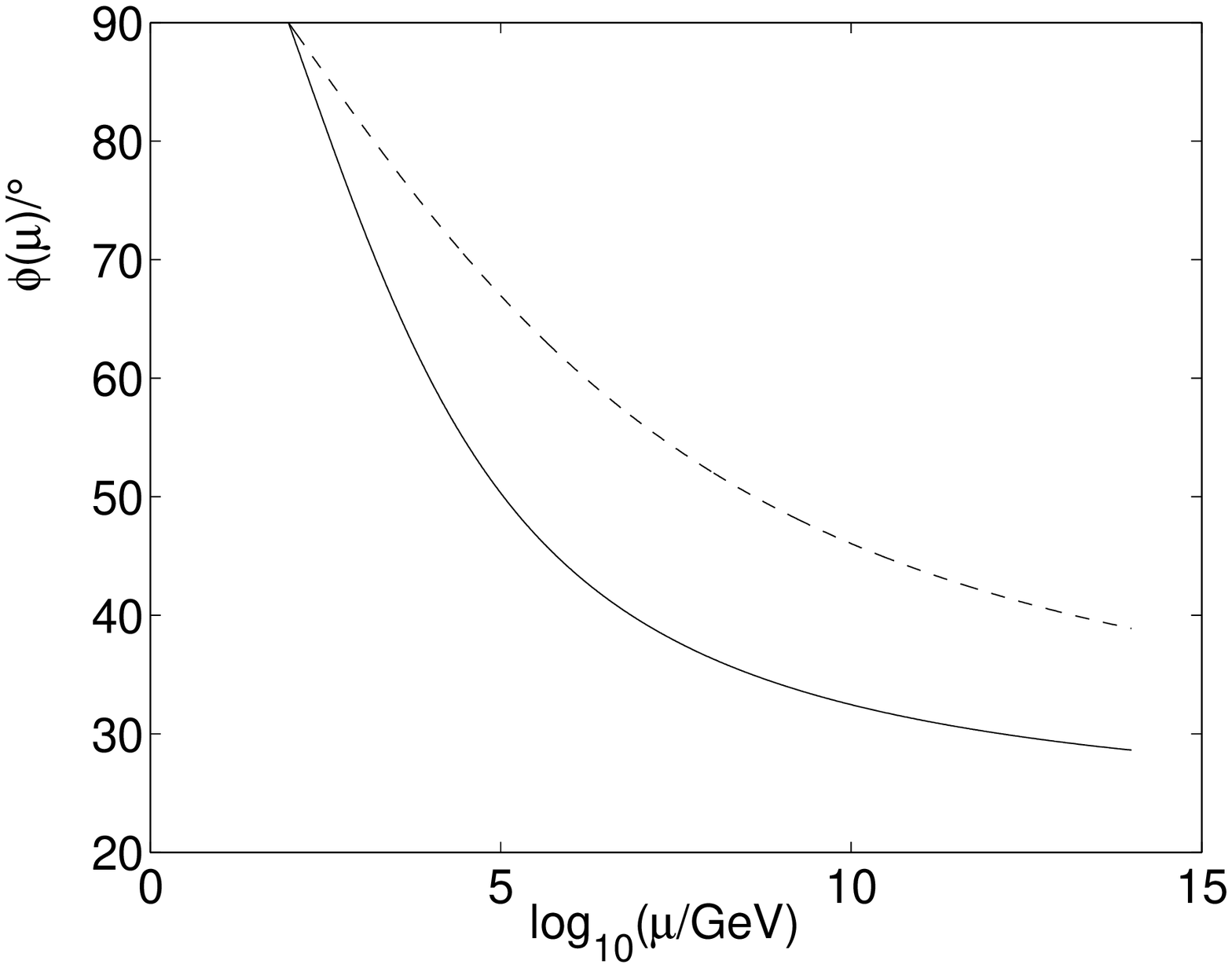,bbllx=-19cm,bblly=18.5cm,bburx=-15cm,bbury=22.5cm,%
width=1.5cm,height=1.5cm,angle=0,clip=0}\vspace{5.3cm}
\caption{The running behaviors of $\theta^{}_l$, $\theta^{}_\nu$,
$\theta$ and $\phi$ in the ND case with $\Delta m^2_{32}
>0$, $\tan\beta=50$ and $m^{}_1 (M^{}_Z) =0.2$ eV within the MSSM,
where the dashed and solid curves stand respectively for the Dirac
and Majorana cases.}
\end{figure}

\begin{figure}[t]
\vspace{0cm}
\epsfig{file=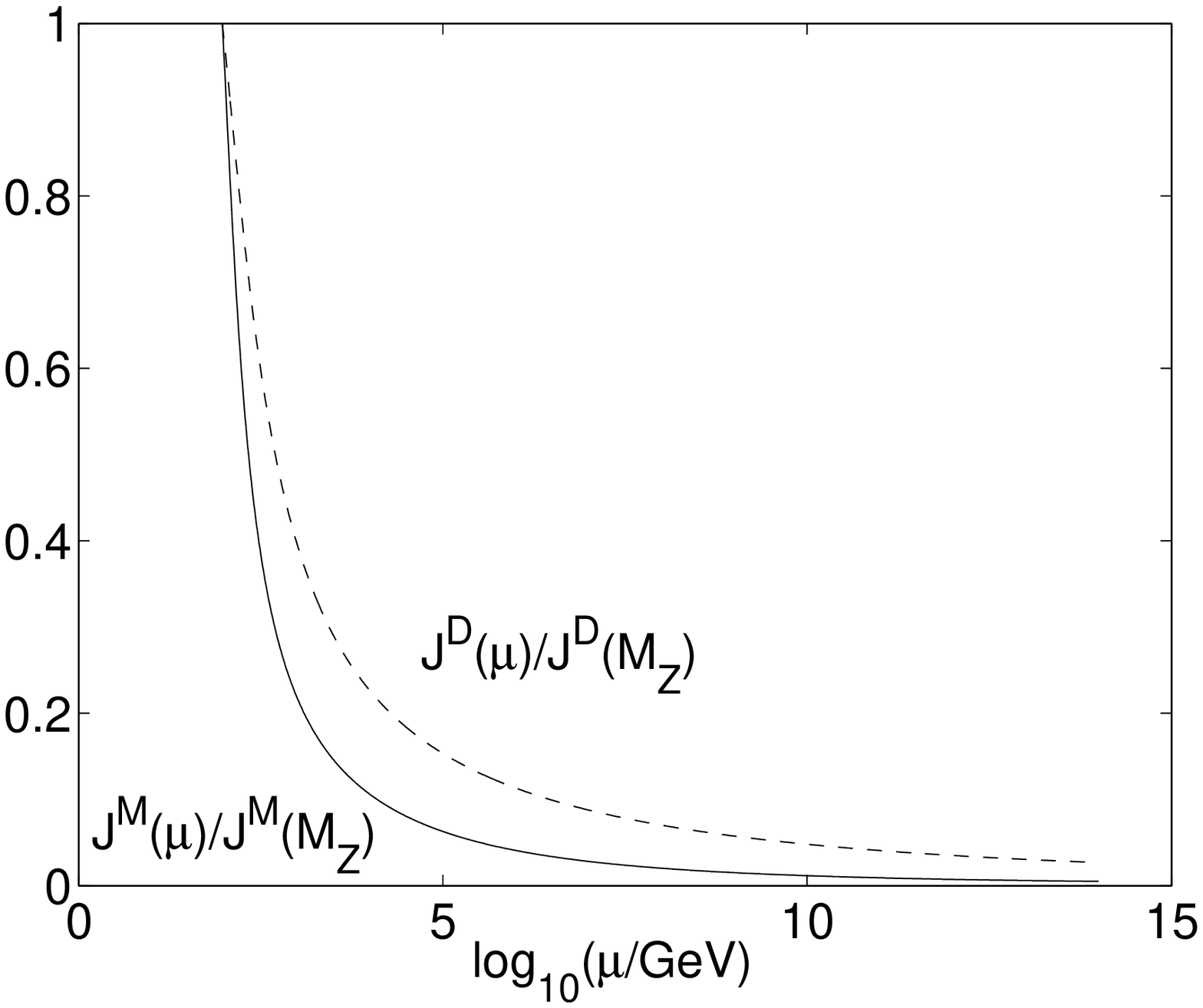,bbllx=-9cm,bblly=18.5cm,bburx=-5cm,bbury=22.5cm,%
width=1.5cm,height=1.5cm,angle=0,clip=0}\vspace{5.1cm}
\caption{The running behavior of $\cal J$ in the ND case with
$\Delta m^2_{32} >0$, $\tan\beta=50$ and $m^{}_1 (M^{}_Z) =0.2$ eV
within the MSSM, where the dashed and solid curves stand
respectively for the Dirac and Majorana cases, and ${\cal J}^{\rm
D} (M^{}_Z) = {\cal J}^{\rm M} (M^{}_Z) \approx 0.0014$.}
\end{figure}

\begin{figure}[t]
\vspace{0cm}
\epsfig{file=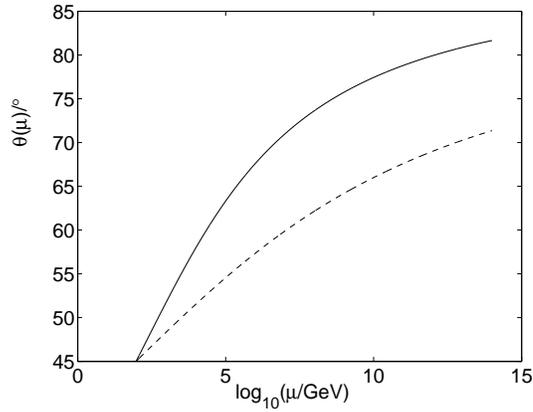,bbllx=-9cm,bblly=18.5cm,bburx=-5cm,bbury=22.5cm,%
width=1.5cm,height=1.5cm,angle=0,clip=0}\vspace{5.1cm}
\caption{The running behavior of $\theta$ in the ND case with
$\Delta m^2_{32} <0$, $\tan\beta=50$ and $m^{}_1 (M^{}_Z) =0.2$ eV
within the MSSM, where the dashed and solid curves stand
respectively for the Dirac and Majorana cases.}
\end{figure}

\end{document}